\documentclass[aps,floatfix]{revtex4}

\usepackage{graphicx}
\usepackage{dcolumn}
\usepackage{bm}
\usepackage{epstopdf}
\usepackage{color}
\usepackage{amssymb}
\usepackage{amsfonts}
\usepackage{xspace} 

\begin{document}

\title[]{High-order harmonic transient grating spectroscopy of SF$_6$ molecular vibrations}

\author{Am\'elie Ferr\'e$^1$, David Staedter$^2$, Fr\'ed\'eric Burgy$^1$, Michal Dagan$^3$, Dominique Descamps$^1$, Nirit Dudovich$^3$, St\'ephane Petit$^1$, Hadas Soifer$^3$, Val\'erie Blanchet$^1$, and Yann Mairesse$^1$}

\address{
$^1$Universit\'e de Bordeaux - CNRS - CEA, CELIA, UMR5107, F33405 Talence, France\\
$^2$Universit\'e de Toulouse - CNRS, LCAR-IRSAMC, Toulouse, France \\
$^3$Department of Physics of Complex Systems, Weizmann Institute of Science, Rehovot 76100, Israel}


\begin{abstract}
Strong field transient grating spectroscopy has shown to be a very versatile tool in time-resolved molecular spectroscopy. Here we use this technique to investigate the high-order harmonic generation from SF$_6$ molecules vibrationally excited by impulsive stimulated Raman scattering. Transient grating spectroscopy enables us to reveal clear modulations of the harmonic emission. This heterodyne detection shows that the harmonic emission generated between 14 to 26 eV is mainly sensitive to two among the three active Raman modes in SF$_6$, i.e. the strongest and fully symmetric $\upsilon_1$-A$_{1g}$ mode (774 cm$^{-1}$, 43 fs) and the slowest mode $\upsilon_5$-T$_{2g}$ (524 cm$^{-1}$, 63 fs). A time-frequency analysis of the harmonic emission reveals  additional dynamics: the strength and central frequency of the $\upsilon_1$ mode oscillate with a frequency of 52 cm$^{-1}$ (640 fs). This could be a signature of the vibration of dimers in the generating medium. Harmonic 11 shows a remarkable behavior, 
oscillating in opposite phase, both on the fast (774 cm$^{-1}$) and slow (52 cm$^{-1}$) timescales, which indicates a strong modulation of the recombination matrix element as a function of the nuclear geometry. These results demonstrate that the high sensitivity of high-order harmonic generation to molecular vibrations, associated to the high sensitivity of transient grating spectroscopy, make their combination a unique tool to probe vibrational dynamics. 
\end{abstract}


\maketitle

\section{Introduction}
The process of high-order harmonic generation (HHG) is being used in a growing number of time-resolved spectroscopy experiment to reveal rotational \cite{itatani2005}, vibrational \cite{Li2008} and electronic dynamics\cite{worner2010, Ruf2012}. While high-harmonic spectroscopy (HHS) is remarkably sensitive, understanding the essence of this sensitivity is a challenging task \cite{Ruf2012,Le2012, walters2009, smirnova2009, spanner2012}. HHG is a highly non-linear process that can be described in three main steps \cite{Corkum93,Krause92}. First, an intense femtosecond laser pulse extracts an electron wave packet by tunnel-ionization from different molecular orbitals. This electron wave packet is then accelerated and coherently scattered in the ionic potential due to the intense electric field of the laser. Concurrently, possible nuclear and electronic dynamics of the cation can take place, depending on the molecular orbitals at play \cite{baker06,smirnova2009, mairesse10}. The last step is the recombination of 
this coherent electron with its cation, resulting in a coherent extreme ultra-violet (XUV) photon emission. 

When HHG is used as a probe of nuclear or electronic dynamics -- instead of more standard techniques like photoionization -- an important question is which of the three steps is/are nuclear or electronic dependent, and where does the high sensitivity of high-harmonic spectroscopy arise from. This point has been the subject of debates in the past few years, even for simple dynamics such as the linear vibrations of the N$_2$O$_4$ van der Waals molecules \cite{Li2008}. Here, large amplitude molecular vibrations of N$_2$O$_4$ induce significant modulations of the tunnel-ionization \cite{spanner2012} and recombination \cite{Le2012,spanner2012} cross sections, which can be accompanied by switches between ionization channels \cite{Li2008}. As a result the harmonic emission shows deep modulations (30 $\%$) as a function of the N-N internuclear distance.

The high sensitivity of HHG to molecular vibrations is not specific to the case of large one-dimension motion. Indeed, complex vibrational dynamics of SF$_6$ were investigated by high-order harmonic spectroscopy in 2006 by Wagner \textit{et al.} \cite{wagner2006}, 
using an impulsive stimulated Raman excitation (ISR) in a colinear pump-probe setup. Among the six normal modes of SF$_6$, three are Raman active: $\upsilon_1$-fully symmetric and strongly active mode $A_{1g}$ with a quantum of 774 cm$^{-1}$ (vibrational period of $\approx$ 43 fs), $\upsilon_2$ the doubly degenerated mode $E_g$ with a quantum of 643 cm$^{-1}$ ($\approx$ 52 fs) and finally the triply degenerated $\upsilon_5$-$T_{2g}$ mode with a quantum of 524 cm$^{-1}$ ($\approx$ 63 fs) \cite{Boudon2004}. ISR is quite inefficient in SF$_6$ with relatively long (25-30 fs) pump pulses, so that only the first vibrational excited state was populated in \cite{wagner2006}, resulting in very weak distorsions of the molecular bond geometry as a function of pump-probe delay. In spite of this, significant modulations of the high-order harmonic emission were measured. All harmonics detected between 35 and 73 eV were found to oscillate at the frequencies of all three Raman modes with a dominant contribution from the 
slowest mode $\upsilon_5$ \cite{wagner2006}. A decay of this mode with a $\sim 1$ ps time constant was assigned to the relaxation of the anisotropy of excitation induced by the ISR. Walters \textit{et al.} have simulated this experiment by sophisticated calculations \cite{walters2009}. In their model, HHG takes place from the ground electronic state of the cation defined by a Jahn-Teller coupling. Extra Raman transitions from the probe pulse and the cross-terms between the vibrational levels of the cation are implemented both in the ionization and recombination steps.  The harmonic emission is defined as the result from the interference of the ground and excited vibrational states. Despite their complexity, these calculations do not quantitatively reproduce the experimental data. 

Here we present an investigation of the XUV emission from vibrationnally excited SF$_6$ molecules in an energy range univestigated before, from 14 to 26 eV, namely between the 9th and the 17th harmonic of a 800 nm femtosecond probe pulse. The dynamic range accessed in colinear pump-probe high-order harmonic spectroscopy, as used by Wagner \textit{et al.}, can be quite low leading to unfavorable signal-to-noise (S/N) ratios, especially for low harmonics which are expected to be weakly modulated \cite{wagner2006}. A detection based on a lock-In amplification to improve the S/N ratio is moreover prevented by the 20 Hz repetition rate of the CCD camera that records the XUV spectrum. This issue can be circumvented by transient grating spectroscopy (TGS), an elegant technique for measuring ultrafast dynamics in solids, 
liquids, or gases, whenever background suppression is required. This femtosecond degenerate wave mixing technique is broadly used in conventional perturbative nonlinear spectroscopy and has been recently extended to nonperturbative nonlinear optics like HHG \cite{Mairesse2008}. It is now used to investigate both the process of HHG itself and ultrafast dynamics like rotational wave packets \cite{Mairesse2008}, vibronic couplings \cite{Ruf2012} or photodissociation \cite{worner2010}, using HHG as a spectroscopic tool. Experimentally, two non-colinear pump pulses create a grating of excitation in the gas jet, through which the intense probe pulse generates high-order harmonics.

The remainder of the paper starts with a description of the experimental setup in section 2. Next, the modulation of the harmonic emission by the different vibrational modes is discussed in section 3. Finally, in section 4, a time-frequency analysis of the signal is performed, revealing unexpected slow oscillations which could be due to SF$_6$ dimer vibrations.

\section{Experimental Set-up} \label{sec:setup}
The experiment was performed using the Aurore laser system at CELIA, which delivers 25 fs, 7 mJ, 800 nm pulses at 1kHz. Half of the beam energy is used for the probe beam, sent through a computer-controlled delay stage towards the high-order harmonic generation chamber. The other half is further split into two by a 50:50 beam splitter, in order to generate two equally intense 800 nm pump pulses. These two beams are then aligned parallel to each other with a vertical offset of $\sim$3 cm, and combined with the third probe beam via a mirror at 45$^{\circ}$ before the vacuum chamber. The probe beam passes through a hole in this mirror so that afterwards all three beams travel in the same vertical plane with an vertical offset between them. The three beams are focused via a 500 mm lens in a continuous molecular jet produced by a 60 $\mu$m nozzle backed by 1 bar of pure SF$_6$. The spatial fringe spacing of the transient grating is 18 $\mu$m and with a waist size of 100 $\mu$m of the probe beam that generates 
high-order harmonics, the diffraction occurs through five fringes. The polarizations of the pump and probe laser beams are parallel to each other. In order to analyze the XUV spectrum, the generated high-order harmonics are sent to an XUV spectrometer consisting of a grating with a groove spacing of 1200 mm$^{-1}$ that images the XUV radiation onto a detector, which consists of a set of dual MCP's, a phosphor and a 20 Hz-CCD camera.

The sinusoidal spatial modulation of the electromagnetic field across the molecular beam resulting from the optical interferences between the two pump beams, leads to a sinusoidal variation of excited and unexcited molecules. As amplitude and phase of the generated high harmonics are different for vibrationally excited and unexcited molecules in the near field, the grating leads to a partial diffraction of the high harmonic emission generated from the probe pulse in the far field. 

Typically energies of 2$\times$ 140 $\mu$J/p and 320 $\mu$J/p were used for the pump and probe pulse, respectively. Such high pump intensities are difficult to use in co-linear geometry because so intense beams can produce themselves high harmonics, resulting in a background signal and an inevitable enhancement of the noise level. 
Using too high pump intensity can result as well in a grating of free electrons due to ionization. This grating of ionization would not evolve on a femtosecond timescale but would produce permanent diffraction peaks in the far field, with which the signal from the molecular excitation grating would optically interfere. While such an grating of ionization does not preclude observing the dynamics, it prevents the extraction of the harmonic phase from the diffraction pattern \cite{mairesse2010}. 
 \begin{figure}[ht!]
 \centering
   \includegraphics[width=0.8\textwidth]{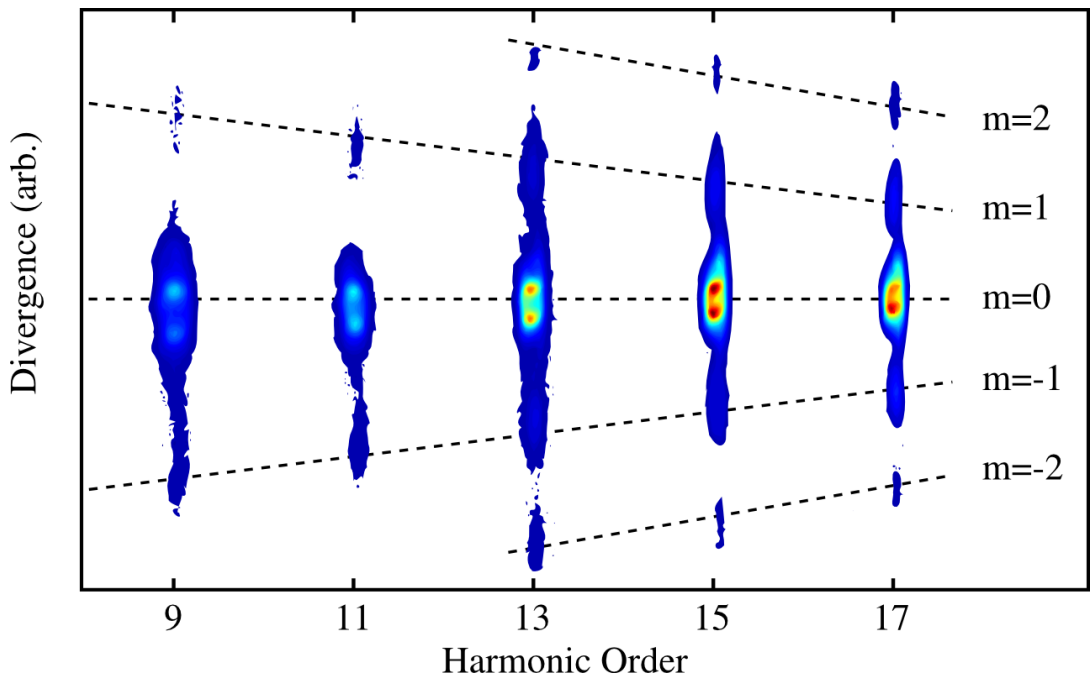}
   \caption{Spatially resolved harmonic spectrum, showing first and second order diffraction peaks. The spatial modulations of the undiffracted light (m=0) result from damages on the phosphor screen. The pump-probe delay is 300 fs.}
   \label{fig1} 
 \end{figure}  
Figure \ref{fig1} shows a recorded image at a pump-probe delay of 300 fs. The fringes of the excitation grating being horizontal in the generating medium, the diffraction peaks appear above and below the main harmonic beam in the far field. Zero (m=0), first (m=$\pm$1) and second (m=$\pm$2) diffraction orders are visible. The diffraction angle $\alpha_q^m$, given by $\alpha_q^m\propto m\times\frac{\lambda_{probe}}{q\lambda_{pump}}$, is twice larger for the second order compared to the first one. Consequently, the second order of diffraction for H(q=9) and H(q=11) are not visible on our detector. Note that with an ionization potential of SF$_6$ at 15.7 eV, the 9th harmonic should result from a multiphoton process, for which the response to molecular dynamics can be very different from that of higher harmonics \cite{soifer10}.

\section{Analysis of the vibrational modes}

\begin{figure}[ht!]
 \centering
   \includegraphics[width=0.8\textwidth]{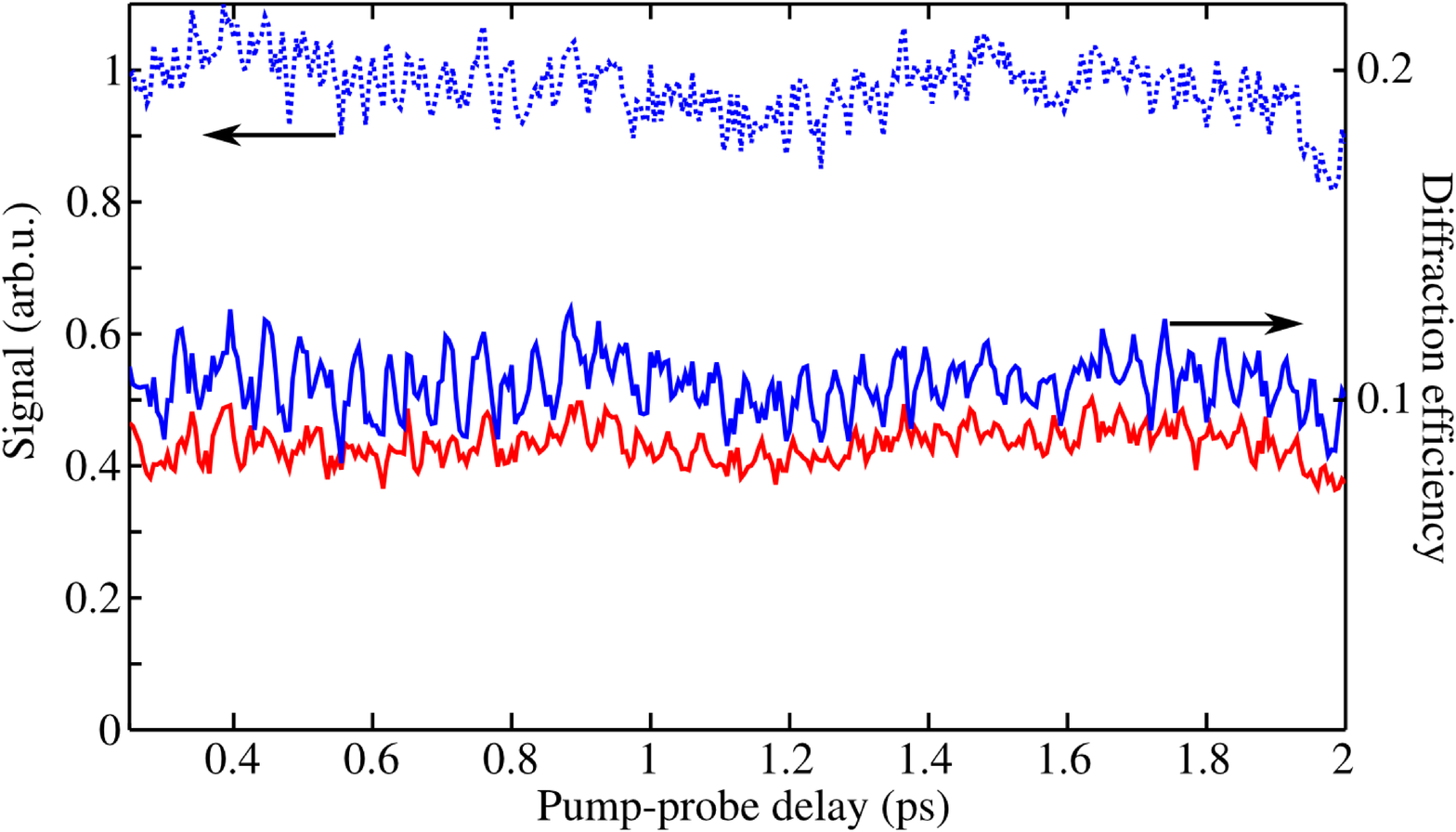}
   \caption{Total harmonic signal (dots) as a function of pump probe delay,
and diffraction efficiency (continuous) for harmonic 11 (red) and 13
(blue). The oscillations of the diffraction efficiency are in opposite
phase for these two harmonics.}
   \label{fig2} 
 \end{figure}

Figure \ref{fig2} shows the evolution of the spatially integrated harmonic signal as a function of the pump-probe delay. The dotted line is the total signal of harmonic 13 ($I^{q=13}_{tot}$). The data is relatively noisy and its Fourier transform does not reveal the three Raman active vibrational modes. This is mostly due to the high intensity of the pump pulses. Despite the apparent lack of oscillations of the total harmonic signal, transient grating spectroscopy is able to reveal the modulations of the harmonic emission in the diffracted light. The first order diffraction efficiency for harmonic $q$ is defined as $\eta^q=\frac{1}{2}\left(I^q_{m=1}+I^q_{m=-1}\right)/I^q_{tot}$, where $I^q_{m=\pm 1}$ is the intensity of the signal diffracted in order $m=\pm 1$ and $I^q_{tot}$ is the total harmonic signal. This diffraction efficiency shows a clear oscillation pattern as a function of pump-probe delay, as illustrated for H11 and H13 in Figure \ref{fig2}. The diffraction efficiency is typically around 10$\%$ 
for each harmonic with an oscillating amplitude increasing from 10 to 30\% from low to high harmonics as already mentionned. We believe that given the high intensity of the pump pulses, part of this diffraction results from a permanent grating of ionization. 

\begin{figure}[ht!]
 \centering
   \includegraphics[width=0.8\textwidth]{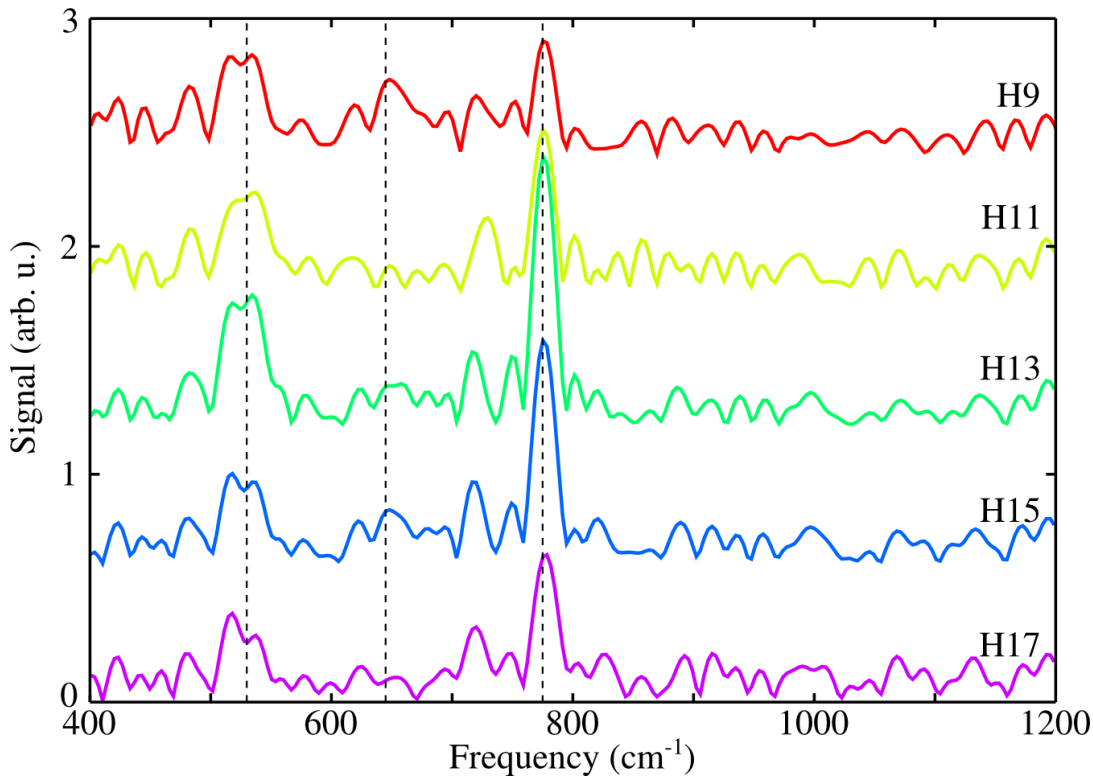}
   \caption{Fast fourier transform of the first order diffraction efficiency, for
harmonic 9 (top) to 17 (bottom). The vertical dashed lines indicate the
frequency of the three excited Raman modes of SF$_6$: 524, 643 and 774
cm$^{-1}$}
   \label{fig3} 
 \end{figure}
\begin{figure}[ht!]
 \centering
   \includegraphics[width=0.8\textwidth]{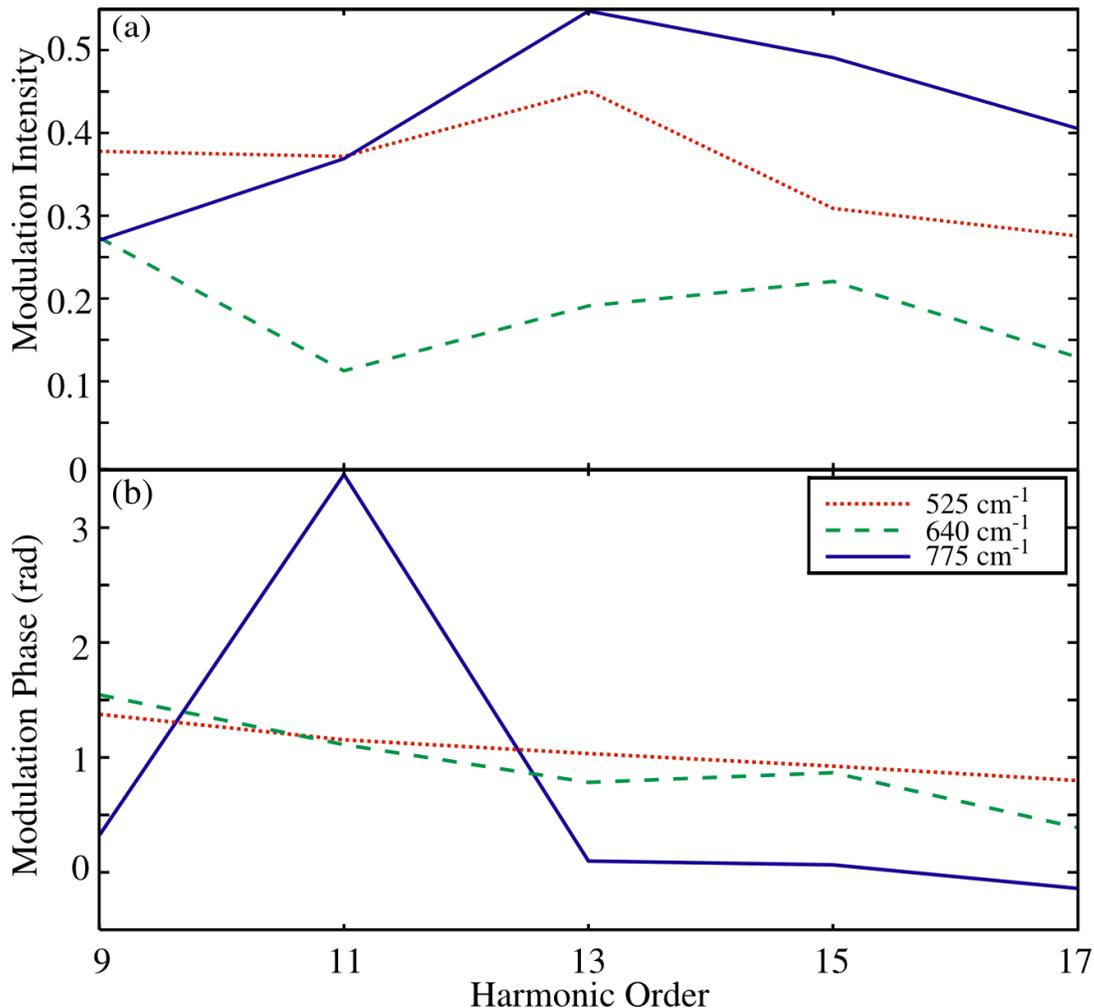}
   \caption{(a) Relative intensity of the different peaks in the harmonic modulation spectrum. The modulation intensity is obtained by integrating each peak in the FFT and normalizing by the total sum of the FFT. (b) Phase of the modulation of the diffraction efficiency. Harmonic 11 oscillates in opposite phase ($\pi$ phase shift) with respect to other harmonics in the dominant 774 cm $^{-1}$ mode. The two other modes produce modulations with the same phase, shifted by $\pi/4$ with respect to the dominant mode. A linear dephasing of the modulation with respect to harmonic order is noticeable.}
   \label{fig4} 
 \end{figure}

 
Figure \ref{fig3} shows the Fourier transform of the diffraction efficiency where a clear modulation of the high harmonic emission is visible, due to the three Raman active modes. The intensity of the peaks in the Fourier transfom, normalized by the total signal in the Fourier transform, are shown in Figure \ref{fig4}(a) as a function of harmonic order. The observed relative weights between these three modes are different from those obtained in the previous (colinear) studies \cite{wagner2006}. Wagner \textit{et al.} found that the slowest mode $\upsilon_5$-T$_{2g}$ (524 cm$^{-1}$, 63 fs) was the dominant one \cite{wagner2006}. In our results the amplitude of $\upsilon_1$-A$_{1g}$ mode (774 cm$^{-1}$, 43 fs) is slightly larger than the ones of $\upsilon_5$, except for H9. The 643 cm$^{-1}$ vibrational mode is above the $\sim$0.15 noise level only for harmonic 9, 13 and 15. Harmonic 13 has the largest oscillations in both modes $\upsilon_1$ and $\upsilon_5$, although H15 and H17 are more intense. More 
quantitatively, Wagner \textit{et al.} measured ratios varying for H39 between $75\%/8\%/17\%$\cite{wagner2006} and $48\%/10\%/42\%$ \cite{walters2007} for the 524/643/774 cm$^{-1}$ modes whereas our results give $41\%/30\%/29\%$ for H9 and $35\%/15\%/50\%$ for H17. Note that the 643 cm$^{-1}$ mode is probably overestimated because of the noise level. 

From the excitation point of view, the fully symmetric $\upsilon_1$-A$_{1g}$ is known as the strongest Raman active mode with an amplitude 20 times larger than the other modes \cite{wagner2006}. The importance of the other modes in the high harmonic modulation is thus striking, indicating a very high sensitivity to some specific molecular distorsions. This question was theoretically investigated by Walters \textit{et al.}, who calculated the HHG from Raman-excited SF$_6$ molecules, taking only a single electronic channel, the highest occupied molecular orbital (HOMO), into account \cite{walters2007,walters2009}. They interpreted the modulations of the harmonic signal in terms of interference from different vibrational channels. They still obtained the $\upsilon_1$ mode as the main mode, like in conventional Raman spectroscopy, with the following relative weights for the modulations of the harmonic emission: $5.6\%/7\%/87\%$ for the 524/643/774 cm$^{-1}$ modes, respectively. These weights do not vary 
drastically when higher vibrational states are considered (up to $\upsilon_i$=4 in each mode). 

Due to its high degree of symmetry (O$_h$) and the hypercoordination of the sulfur atom, five different states of the cation are lying within only a 10 eV range above the ionization threshold of SF$_6$ \cite{Holland92}. Indeed, we have recently shown that high harmonic emission from SF$_6$ involves many possible ionization channels and that the dominant channel depends on the energy range \cite{FerreSF6}. We expect that the high harmonic response to molecular vibrations will be different for these different channels. The difference between our experimental results and the ones from Wagner \textit{et al} could be thus  caused by the different harmonic orders investigated. This could also explain why the measurements from Wagner \textit{et al.} disagree with the theoretical study from Walters \textit{et al.} who only considered the lowest cation electronic state in their calculation. The better agreement with our results to the calculations, suggests that the ionic states involved in the generation of low 
harmonics 
gives a response to vibrations close to the response of the highest occupied molecular orbital. Note that there are other sources of differences between our experiment and the previous ones. First, we use a higher (three times) pump intensity. Second, our detection is more sensitive to phase modulations of the harmonic emission \cite{mairesse2010}. And last but not least, the scanned pump-probe delay range is twice longer in our case. As already noted by Wagner et al., the $\upsilon_5$ vibrational quantum beats disappear after $\approx $1 ps, changing significantly the Fourier spectrum. Consequently, a simple Fourier Transform of the pump-probe signal does not deliver the complete image of the undergoing dynamics. A deeper analysis will be presented in the next section. 

Additional information on the high harmonic response to molecular vibration can be obtained by studying the phase of the modulation for each mode, extracted by the Fourier transform. The most striking feature in Figure \ref{fig4} (b), which shows the phase of the modulation of the diffraction efficiency, is a $\pi$ phase shift between the oscillations of H11 and the neighboring harmonics for the 774 cm $^{-1}$ mode. This completely out of phase time dependency is clearly visible on the raw data presented in Figure \ref{fig2} for delays larger than 1 ps. The modulation phase also shows a continuous decrease with harmonic order, and a $\sim\pi/4$ shift between the 774 cm $^{-1}$ mode and the two other modes. Note that the duration of the pump and probe pulses being both $\approx$ 30 fs, significant nuclear dynamics take place during both the pump and the probe interaction. Consequently, the determination of the zero delay can not be achieved precisely and the phases extracted from the Fourier transform are not 
absolute. Only relative phases are meaningful.

The phase of the oscillations reflects the evolution of the harmonic signal with molecular stretching. If the pump pulse was much shorter than the vibrational period $T$, the molecule could be considered frozen during the pump interaction. An efficient ISR excitation would then create a vibrational wave packet initially localized at the equilibrium geometry. Assuming a positive gradient of the polarizability, the early nuclear dynamics would tend to a stretched geometry, maximizing at $T/4$ and periodically every $T/4+kT$, $k$ being an integer. Therefore, if the harmonic intensity or phase is maximized for such a fully stretched geometry, the diffraction efficiency will oscillate with maxima at $T/4+kT$. By contrast, if the optimal geometry for diffraction efficiency is reached on the inner part of the molecular energy potential, the oscillations will be shifted by $\pi$. The $\pi$ shifted modulation phase of H11 for the  774 cm $^{-1}$ mode thus indicates an inverted sensitivity of the harmonic amplitude or 
phase to the molecular stretching in this particular mode. More generally, the strong variation of the phase with respect to the vibrational mode indicates that the HHG process in this low energy range depends strongly on the different geometries scanned by the vibrational coherences. Further experimental investigations at higher harmonic orders are necessary to conclude if this sensitivity is enhanced by the transient grating detection or simply due to the investigated energy range, which was lower than in the previous study \cite{wagner2006}. 

\section{Revealing the vibrations of dimers}  
 \begin{figure}[ht!]
 \centering
   \includegraphics[width=0.8\textwidth]{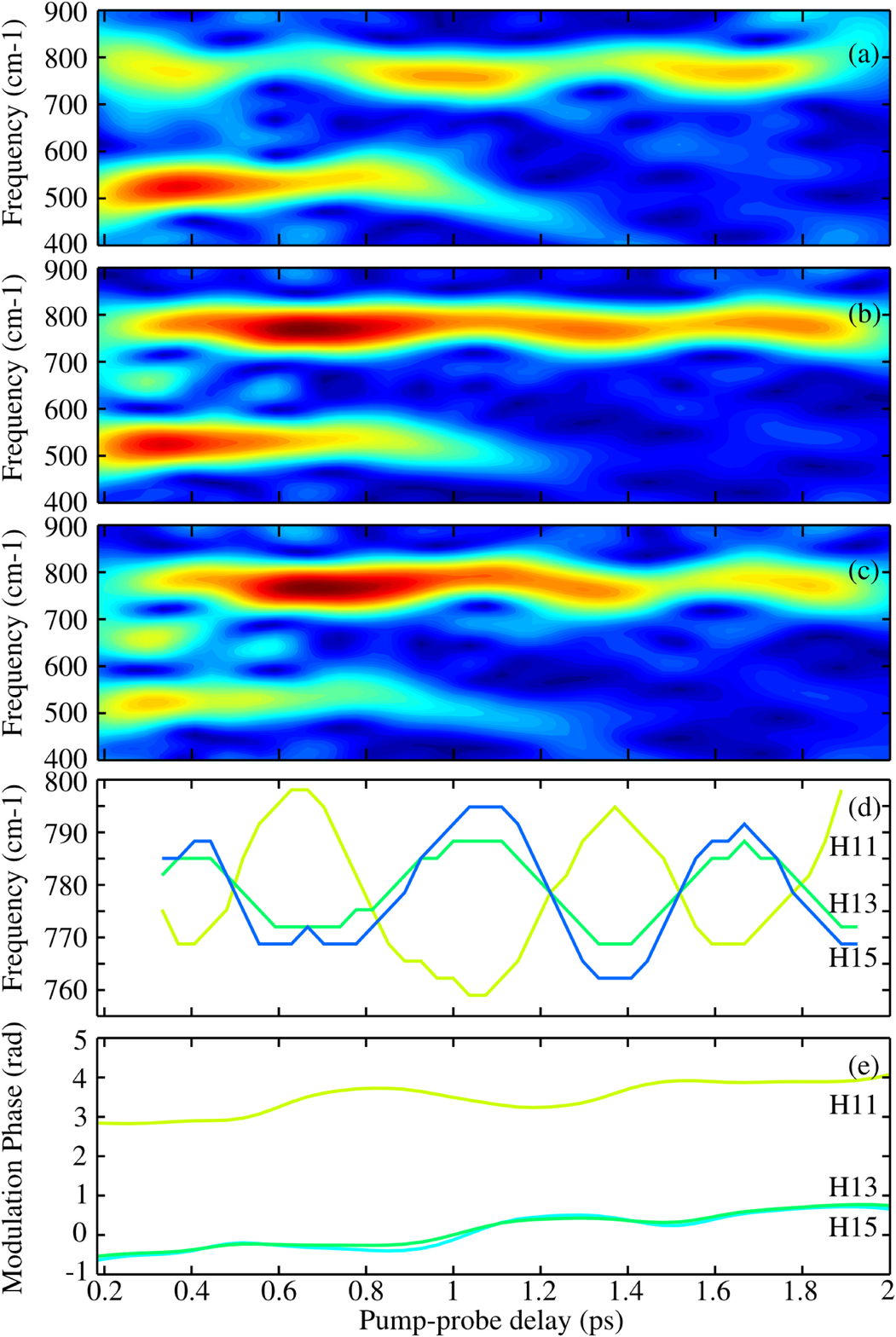}
   \caption{Gabor analysis using a 400 fs FWHM  sliding FFT, to get the time-frequency distribution of the diffraction efficiency for
harmonic 11 (a), 13 (b) and 15 (c). Central frequency (d) and phase (e) of the peak corresponding to the $\upsilon_1$ mode, for harmonic 11 to 15 as a function of pump probe delay. }
   \label{fig5} 
 \end{figure}
Up to now the whole pump-probe scan was used in the Fourier analysis, which means that the modulation of the amplitude and phase were averaged over 2 ps. However, on the raw diffraction efficiency a slow evolution of the oscillation pattern appears in Figure \ref{fig2}. The oscillations for H13 are better defined for delays above 1 ps. In order to understand the origin of this effect, we conducted a time-frequency analysis of the diffraction efficiency. We used a sliding supergaussian window function with 400 fs full-width at half maximum. The resulting spectrograms are shown in Figure \ref{fig5}. The weak contribution observed on $\upsilon_2$ around 650 cm$^{-1}$ in Figure \ref{fig3} results in fact from a very fast damping of this mode in less than 300 fs, as displayed in Figure \ref{fig5}. Similarly, an exponential decay of the asymmetric mode at 524 cm$^{-1}$ is clearly observed with a typical timescale of 930 $\pm 150$ fs, independently of the harmonic order. This decay explains the better apparent 
contrast of the oscillations in the raw signal of Fig. \ref{fig2} once the mode is off. The fully symmetric mode shows an intriguing behavior: the intensity of the peak as well as its central frequency oscillate around 774 cm$^{-1}$ with a period of 52 cm$^{-1}$. The central frequency oscillates over a 40 cm$^{-1}$ range, as shown in Figure \ref{fig5}(d). Remarkably, H11 is once more completely out of phase with respect to H13: the peak for H11 is maximized and redshifted when the one for H13 is minimized and blueshifted. This slow oscillation, and the specific behavior of H11, have been systematically observed at several pump energies. The range of the frequency oscillation tends to increase when the pump pulse intensity is increased. A similar tendency is observed on the 524 cm$^{-1}$ mode but with the same temporal evolution on all the harmonic orders. Last, we investigated the phase of the harmonic modulation at 774 cm$^{-1}$ as a function of pump-probe delay, to determine whether or not the opposite 
phase of harmonic 11 observed in Fig. \ref{fig4} persists throughout the delay scan. The results  (Fig. \ref{fig5}(e)) show that the modulation phase of the different harmonics slowly oscillate in time, as the result of the modifications of the central frequency of the oscillations (the phase is evaluated at a fixed frequency). The modulation phase of harmonic 11 remains strongly shifted with respect to higher harmonics, by 2.8 to rad depending on the delay. 

What could be the origin of the slow oscillations of the $\upsilon_1$ mode, both in amplitude and central frequency? ISR is so inefficient in SF$_6$ that only a few vibration quanta can be coherently populated by the pump pulses. The high splitting in energy cannot result from vibrational anharmonicities or band combination, which are in the range of $\approx$1 cm$^{-1}$ \cite{McDowell86,McDowell86bis}. The rotational temperature cannot be responsible for such shifts, since assuming a rotational temperature of 300 K leads to a main rotational energy of $\approx$6 cm$^{-1}$. The ground electronic state being singlet, no spin-orbit coupling is expected. The first excited state is laying above 6 eV, inhibiting efficient HHG due to the reduced 
ionization threshold. In conclusion, such frequency beating can not be explained by the spectroscopy of SF$_6$ monomer. However, nucleation of SF$_6$ is possible in our molecular beam conditions \cite{GERAEDTS81,okada}, leading to a significant proportion of dimers or even larger clusters in the beam. 

Theoretical studies of the SF$_6$ dimers have been conducted, giving a bound energy around 450-600 cm$^{-1}$ and an internuclear distance of about 4.9 Angstrom \cite{bladel,beu95}. The calculated geometrical structure has a D$_{2d}$ symmetry with four S-F axes aligned along the S$_6$ axis of the pseudo-rotation of the monomer, due to the electronegativity of the fluor. This van der Waals dimer has consequently four vibrational modes at frequencies smaller than 40 cm$^{-1}$, all of them being Raman active. The main experimental signature of the dimer has been observed in IR spectroscopy as a 20 cm$^{-1}$ splitting of the $\upsilon_3$ at 948 cm$^{-1}$, due mainly to a dipole-dipole interaction \cite{urban}. 

No Raman spectroscopy has been achieved on SF$_6$ dimers to the best of our knowledge. We can assume that the SF$_6$ moieties in the dimer will oscillate around the frequency 774 cm$^{-1}$ due to a Raman process induced by the pump pulse. However, being in an induced-dipole interaction with another SF$_6$ moiety, this rapid $\upsilon_1$ motion will be altered by the slow variation of the geometry of the dimer, causing a slow variation of the central frequency. Indeed, the dimer itself is Raman activated with for instance a torsion mode calculated at 22 cm$^{-1}$ or a stretching mode calculated at 30 cm$^{-1}$ \cite{bladel}. The present heterodyne detection on the HHG yield is sensitive to such a coupling and the fact that an out of phase oscillation is observed on H11 reveals that the sensitivity must be mainly established during the recombination step. In conclusion, high harmonic emission from both unperturbed (monomer) and perturbed by a coherent surrounding (dimer) SF$_6$ takes place, resulting in a 
complex quantum beating.

\section{Conclusion}
We have shown that high harmonic transient grating spectroscopy is remarkably sensitive to molecular vibrations. Our study confirms that modes which are very weakly Raman-active modulate the harmonic emission with high contrast. The weight of the different modes in the harmonic modulation differs from previous experimental studies, probably because of the different spectral energy range investigated which implies that different ionization channels are involved. The heterodyne nature of our detection scheme, which is more sensitive to phase modulations, could also be a source of differences. Both the relative weight of the modes and their phases are found to vary with harmonic order. In particular, the modulations of the diffraction efficiency for H11 in the $\upsilon_1$ (774 cm$^{-1}$) mode are out of phase with the other harmonics, which is the signature of an inverted dependence of the harmonic intensity or phase with the stretch coordinate. This probably results from a sharp structure in the recombination 
matrix elements, such as an autoionizing resonance. 

The time-frequency analysis of the diffraction efficiency reveals slow oscillations of the modulation amplitude and central frequency around the $\upsilon_1$ mode, with a $\sim 50$ cm$^{-1}$ frequency. These slow oscillations are in opposite phase for H11, as are the fast oscillations of the diffracted signal at the 774 cm$^{-1}$ frequency. This unexpected feature could be due to vibrations of the dimer which periodically modulate the distance between the SF$_6$ centers and consequently the associated $\upsilon_1$ mode. Further investigations are required to confirm this interpretation. From an experimental point of view, this study appeals for a Raman spectroscopic study of SF$_6$ dimers. In addition, the measurements could be repeated in heavy clustering conditions \cite{Ruf2013} to monitor the influence of the concentration of dimers on the diffracted harmonic signal.
 
We thank Rodrigue Bouillaud and Laurent Merzeau for technical assistance and Vincent Boudon, Fabrice Catoire and Jean-Christophe Delagnes for scientific discussions. This work was supported financially by the Conseil Regional d'Aquitaine (20091304003 ATTOMOL and COLA 2 No. 2.1.3-09010502). D.S. acknowledges the European Union for its PhD fellowship from the ITN-ICONIC (FP7-Grant Agreement No. 238671).


\begin{thebibliography}{0}
\expandafter\ifx\csname natexlab\endcsname\relax\def\natexlab#1{#1}\fi
\expandafter\ifx\csname bibnamefont\endcsname\relax
  \def\bibnamefont#1{#1}\fi
\expandafter\ifx\csname bibfnamefont\endcsname\relax
  \def\bibfnamefont#1{#1}\fi
\expandafter\ifx\csname citenamefont\endcsname\relax
  \def\citenamefont#1{#1}\fi
\expandafter\ifx\csname url\endcsname\relax
  \def\url#1{\texttt{#1}}\fi
\expandafter\ifx\csname urlprefix\endcsname\relax\def\urlprefix{URL }\fi
\providecommand{\bibinfo}[2]{#2}
\providecommand{\eprint}[2][]{\url{#2}}

\end{thebibliography}


\section*{References}
\begin{thebibliography}{10}

\bibitem{itatani2005}
J.~Itatani, D.~Zeidler, J.~Levesque, Michael Spanner, D.~M. Villeneuve, and
  P.~B. Corkum.
\newblock {\em Phys. Rev. Lett.}, 94(12):4, 2005.

\bibitem{Li2008}
Wen Li, Xibin Zhou, Robynne Lock, Serguei Patchkovskii, Albert Stolow, Henry~C.
  Kapteyn, and Margaret~M. Murnane.
\newblock {\em Science}, 322(5905):1207--1211, 2008.

\bibitem{worner2010}
H.~J. Worner, J.~B. Bertrand, D.~V. Kartashov, P.~B. Corkum, and D.~M.
  Villeneuve.
\newblock {\em Nature}, 466(7306):604--607, 2010.

\bibitem{Ruf2012}
H.~Ruf, C.~Handschin, A.~Ferre, N.~Thire, J.~B. Bertrand, L.~Bonnet,
  R.~Cireasa, E.~Constant, P.~B. Corkum, D.~Descamps, B.~Fabre, P.~Larregaray,
  E.~Mevel, S.~Petit, B.~Pons, D.~Staedter, H.~J. Worner, D.~M. Villeneuve,
  Y.~Mairesse, P.~Halvick, and V.~Blanchet.
\newblock {\em J. Chem. Phys.}, 137(22):224303, 2012.

\bibitem{Le2012}
A.~T. Le, T.~Morishita, R.~R. Lucchese, and C.~D. Lin.
\newblock {\em Phys. Rev. Lett.}, 109(20):5, 2012.

\bibitem{walters2009}
Z.~B. Walters, S.~Tonzani, and C.~H. Greene.
\newblock {\em Chemical Physics}, 366:103--114, 2009.

\bibitem{smirnova2009}
Olga Smirnova, Yann Mairesse, Serguei Patchkovskii, Nirit Dudovich, David
  Villeneuve, Paul Corkum, and Misha~Yu Ivanov.
\newblock {\em Nature}, 460(7258):972--977, 2009.

\bibitem{spanner2012}
Michael Spanner, Jochen Mikosch, Andrey~E. Boguslavskiy, Margaret~M. Murnane,
  Albert Stolow, and Serguei Patchkovskii.
\newblock {\em Phys. Rev. A}, 85(3):033426, 2012.

\bibitem{Corkum93}
P.~B. Corkum.
\newblock Plasma perspective on strong field multiphoton ionization.
\newblock {\em Physical Review Letters}, 71(13):1994--1997, 1993.

\bibitem{Krause92}
Jeffrey~L. Krause, Kenneth~J. Schafer, and Kenneth~C. Kulander.
\newblock High-order harmonic generation from atoms and ions in the high
  intensity regime.
\newblock {\em Physical Review Letters}, 68(24):3535--3538, June 1992.

\bibitem{baker06}
S.~Baker, J.~S. Robinson, C.~A. Haworth, H.~Teng, R.~A. Smith, C.~C. Chirilă,
  M.~Lein, J.~W.~G. Tisch, and J.~P. Marangos.
\newblock Probing proton dynamics in molecules on an attosecond time scale.
\newblock {\em Science}, 312(5772):424--427, April 2006.

\bibitem{mairesse10}
Y.~Mairesse, J.~Higuet, N.~Dudovich, D.~Shafir, B.~Fabre, E.~M{\'e}vel,
  E.~Constant, S.~Patchkovskii, Z.~Walters, M.~Yu. Ivanov, and O.~Smirnova.
\newblock High harmonic spectroscopy of multichannel dynamics in strong-field
  ionization.
\newblock {\em Physical Review Letters}, 104(21), May 2010.

\bibitem{wagner2006}
Nicholas~L. Wagner, Andrea Wuest, Ivan~P. Christov, Tenio Popmintchev, Xibin
  Zhou, Margaret~M. Murnane, and Henry~C. Kapteyn.
\newblock {\em PNAS}, 103(36):13279--13285, 2006.

\bibitem{Boudon2004}
V.~Boudon, J.~L. Dom{\'e}nech, D.~Bermejo, and H.~Willner.
\newblock {\em J. Molec. Spect.}, 228(2):392--400, 2004.

\bibitem{Mairesse2008}
Y.~Mairesse, D.~Zeidler, N.~Dudovich, M.~Spanner, J.~Levesque, D.~M.
  Villeneuve, and P.~B. Corkum.
\newblock {\em Phys. Rev. Lett.}, 100(14):143903, 2008.

\bibitem{mairesse2010}
Y~Mairesse, N~Dudovich, D~Zeidler, M~Spanner, D~M Villeneuve, and P~B Corkum.
\newblock Phase sensitivity of high harmonic transient grating spectroscopy.
\newblock {\em J. Phys. B}, 43(6):065401, 2010.

\bibitem{soifer10}
H.~Soifer, P.~Botheron, D.~Shafir, A.~Diner, O.~Raz, B.~Bruner, Y.~Mairesse,
  B.~Pons, and N.~Dudovich.
\newblock Near-threshold high-order harmonic spectroscopy with aligned
  molecules.
\newblock {\em Physical Review Letters}, 105(14), September 2010.

\bibitem{walters2007}
B.~Walters, Zachary, Stefano Tonzani, and H.~Greene, Chris.
\newblock {\em J. of Phys. B}, 40(18):F277, 2007.

\bibitem{Holland92}
D.~M.~P. Holland, D.~A. Shaw, A.~Hopkirk, M.~A. MacDonald, and S.~M. McSweeney.
\newblock {\em J. Phys. B}, 25(22):4823, 1992.

\bibitem{FerreSF6}
A.~Ferr\'e, B.~Fabre, V.~Blanchet, F.~Burgy, D.~Descamps, N.~Fedorov,
  J.~Gaudin, G.~Geoffroy, S.~Patchkvoskii, S.~Petit, H.~Soifer, N.~Dudovich,
  and Y.~Mairesse.
\newblock {\em submitted}, 2013.

\bibitem{McDowell86}
Robin~S. McDowell, Burton~J. Krohn, Herbert Flicker, and Mariena~C. Vasquez.
\newblock {\em Spect. Acta Part A: Mol. Spect.}, 42(2,3):351--369, 1986.

\bibitem{McDowell86bis}
Robin~S. McDowell and Burton~J. Krohn.
\newblock {\em Spect. Acta Part A: Mol. Spect.}, 42(2,3):371--385, 1986.
\newblock 0584-8539.

\bibitem{GERAEDTS81}
J.~Geraedts, S.~Setiadi, S.~Stolte, and J.~Reuss.
\newblock {\em Chem. Phys. Lett.}, 78(2):277--282, 1981.

\bibitem{okada}
Y.~Okada, K.~Ashimine, and K.~Takeuchi.
\newblock {\em Applied Physics B}, 70(1):117--121, 2000.

\bibitem{bladel}
J.~W.~I. van Bladel and A.~van~der Avoird.
\newblock {\em J. Chem. Phys.}, 92(5):2837--2847, 1990.

\bibitem{beu95}
T.~A. Beu and K.~Takeuchi.
\newblock {\em J. Chem. Phys.}, 103(15):6394--6413, 1995.

\bibitem{urban}
R.~Urban and M.~Takami.
\newblock {\em J. Chem. Phys.}, 103(21):9132--9137, 1995.

\bibitem{Ruf2013}
H.~Ruf, C.~Handschin, R.~Cieasa, N.~Thir\'e, A.~Ferr\'e, S.~Petit, D.~Descamps,
  E.~M\'evel, E.~Constant, V.~Blanchet, B.~Fabre, and Y.~Mairesse.
\newblock {\em Phys. Rev. Lett.}, 110:083902, 2013.

\end{thebibliography}
\end{document}